\documentclass{emulateapj}
\usepackage{apjfonts}
\usepackage{rotating}
\usepackage{multirow}

\newcommand{\pivec}{\mbox{\boldmath $\pi$}}

\lefthead{PARK ET AL.}
\righthead{BINARY LENSING WITH ORBITAL EFFECTS}

\begin{document}
\title{GRAVITATIONAL BINARY-LENS EVENTS WITH PROMINENT EFFECTS OF LENS ORBITAL MOTION}

\author{
H. Park$^{1,34}$, 
A. Udalski$^{2,33}$, 
C. Han$^{1,34,37}$, 
A. Gould$^{3,34}$, 
J.-P. Beaulieu$^{4,35}$,
Y. Tsapras$^{5,6,36}$, \\
and\\
% OGLE --------------------------
M. K. Szyma\'nski$^{2}$, 
M. Kubiak$^{2}$, 
I. Soszy\'nski$^{2}$, 
G. Pietrzy\'nski$^{2,7}$, 
R. Poleski$^{2,3}$, 
K. Ulaczyk$^{2}$, 
P. Pietrukowicz$^{2}$, \\
S. Koz{\l}owski$^{2}$, 
J. Skowron$^{2}$, 
{\L}. Wyrzykowski$^{2,8}$ \\ 
(The OGLE Collaboration),\\
% microFUN ----------------------
J.-Y. Choi$^{1}$, 
D. L. Depoy$^{9}$, 
Subo Dong$^{10}$, 
B. S. Gaudi$^{3}$, 
K.-H. Hwang$^{1}$, 
Y. K. Jung$^{1}$, 
A. Kavka$^{3}$, 
C.-U. Lee$^{11}$, \\
L. A. G. Monard$^{12}$, 
B,-G. Park$^{11}$, 
R. W. Pogge$^{3}$, 
I. Porritt$^{13}$, 
I.-G. Shin$^{1}$, 
J. C. Yee$^{3}$ \\
(The $\mu$FUN Collaboration),\\
% PLANET ------------------------
M. D. Albrow$^{14}$, 
D. P. Bennett$^{15}$, 
J. A. R. Caldwell$^{16}$, 
A. Cassan$^{4}$, 
C. Coutures$^{4}$, 
D. Dominis$^{17}$, 
J. Donatowicz$^{18}$, \\
P. Fouqu{\'e}$^{19}$, 
J. Greenhill$^{20}$, 
M. Huber$^{21}$, 
U. G. J{\o}rgensen$^{22}$, 
S. Kane$^{23}$, 
D. Kubas$^{4}$, 
J. -B. Marquette$^{4}$, 
J. Menzies$^{24}$, \\
C. Pitrou$^{4}$, 
K. R. Pollard$^{14}$, 
K. C. Sahu$^{25}$, 
J. Wambsganss$^{26}$, 
A. Williams$^{27}$, 
M. Zub$^{26}$ \\
(The PLANET Collaboration),\\
% -------- RoboNet --------------
A. Allan$^{28}$, 
D. M. Bramich$^{29}$, 
P. Browne$^{30}$, 
M. Dominik$^{30}$, 
K. Horne$^{30}$, 
M. Hundertmark$^{30}$, 
N. Kains$^{29}$, 
C. Snodgrass$^{31}$, \\
I. A. Steele$^{32}$, 
R. A. Street$^{5}$ \\ 
(The RoboNet Collaboration)
}

\bigskip\bigskip
\affil{$^{1}$Department of Physics, Institute for Astrophysics, Chungbuk National University, Cheongju 371-763, Korea}
\affil{$^{2}$Warsaw University Observatory, Al. Ujazdowskie 4, 00-478 Warszawa, Poland}
\affil{$^{3}$Department of Astronomy, The Ohio State University, 140 West 18th Avenue, Columbus, OH 43210, USA}
\affil{$^{4}$Institut d'Astrophysique de Paris, UMR 7095 CNRS - Universit{\'e} Pierre \& Marie Curie, 98bis Bd Arago, 75014 Paris, France}
\affil{$^{5}$Las Cumbres Observatory Global Telescope Network, 6740 Cortona Drive, Suite 102, Goleta, CA 93117, USA}
\affil{$^{6}$School of Mathematical Sciences, Queen Mary, University of London, Mile End Road, London E1 4NS, UK}
% OGLE --------------------------
\affil{$^{7}$Universidad de Concepci\'{o}n, Departamento de Astronomia, Casilla 160-C, Concepci\'{o}n, Chile}
\affil{$^{8}$Institute of Astronomy, University of Cambridge, Madingley Road, Cambridge CB3 0HA, UK}
% microFUN ----------------------
\affil{$^{9}$Department of Physics and Astronomy, Texas A\&M University, College Station, TX 77843, USA}
\affil{$^{10}$Institute for Advanced Study, Einstein Drive, Princeton, NJ 08540, USA}
\affil{$^{11}$Korea Astronomy and Space Science Institute, 776 Daedukdae-ro, Yuseong-gu, Daejeon 305-348, Korea}
\affil{$^{12}$Klein Karoo Observatory, Calitzdorp, and Bronberg Observatory, Pretoria, South Africa}
\affil{$^{13}$Turitea Observatory, Palmerston North, New Zealand}
% PLANET ------------------------
\affil{$^{14}$Department of Physics and Astronomy, University of Canterbury, Private Bag 4800, Christchurch 8020, New Zealand}
\affil{$^{15}$Department of Physics, University of Notre Dame, 225 Nieuwland Science Hall, Notre Dame, IN 46556-5670, USA}
\affil{$^{16}$McDonald Observatory, University of Texas, Fort Davis, TX 79734, USA}
\affil{$^{17}$Astrophysikalisches Institut Potsdam, An der Sternwarte 16, 14482 Potsdam, Germany}
\affil{$^{18}$Department of Computing, Technical University of Vienna, Wiedner Hauptstrasse 10, A-1040 Vienna, Austria}
\affil{$^{19}$Observatoire Midi-Pyr{\'e}n{\'e}es, Laboratoire d'Astrophysique, UMR 5572, Universit{\'e} Paul Sabatier - Toulouse 3, 14 avenue Edouard Belin, 31400 Toulouse, France}
\affil{$^{20}$School of Mathematics and Physics, University of Tasmania, Private Bag 37, Hobart, TAS 7001, Australia}
\affil{$^{21}$Institute for Astronomy, University of Hawaii, 2680 Woodlawn Drive Honolulu, HI 96822-1839, USA}
\affil{$^{22}$Niels Bohr Institute, Astronomical Observatory, Juliane Maries vej 30, 2100 Copenhagen, Denmark}
\affil{$^{23}$NASA Exoplanet Science Institute, Caltech, MS 100-22, 770 South Wilson Avenue, Pasadena, CA 91125, USA}
\affil{$^{24}$South African Astronomical Observatory, PO Box 9 Observatory 7935, South Africa}
\affil{$^{25}$Space Telescope Science Institute, 3700 San Martin Drive, Baltimore, MD 21218, USA}
\affil{$^{26}$Astronomisches Rechen-Institut (ARI), Zentrum f\"{u}r Astronomie der Universit\"{a}t Heidelberg (ZAH), M\"{o}nchhofstr. 12-14, 69120, Heidelberg, Germany}
\affil{$^{27}$Perth Observatory, Walnut Road, Bickley, Perth 6076, Australia}
% RoboNet -----------------------
\affil{$^{28}$School of Physics, University of Exeter Stocker Road, Exeter, Devon, EX4 4QL, UK}
\affil{$^{29}$European Southern Observatory, Karl-Schwarzschild-Stra{\ss}e 2, 85748 Garching bei M{\"u}nchen, Germany}
\affil{$^{30}$SUPA, University of St. Andrews, School of Physics and Astronomy, North Haugh, St. Andrews, KY16 9SS, UK}
\affil{$^{31}$Max Planck Institute for Solar System Research, Max-Planck-Str. 2, 37191 Katlenburg-Lindau, Germany}
\affil{$^{32}$Astrophysics Research Institute, Liverpool John Moores University, Twelve Quays House, Egerton Wharf, Birkenhead, Wirral., CH41 1LD, UK}
% -------------------------------
\affil{$^{33}$The OGLE Collaboration}
\affil{$^{34}$The $\mu$FUN Collaboration}
\affil{$^{35}$The PLANET Collaboration}
\affil{$^{36}$The RoboNet Collaboration}
\affil{$^{37}$Corresponding author}

\begin{abstract}
Gravitational microlensing events produced by lenses composed of binary masses 
are important because they provide a major channel to determine physical parameters 
of lenses. In this work, we analyze the light curves of two binary-lens events 
OGLE-2006-BLG-277 and OGLE-2012-BLG-0031 for which the light curves exhibit 
strong deviations from standard models. From modeling considering various 
second-order effects, we find that the deviations are mostly explained by the 
effect of the lens orbital motion. We also find that lens parallax effects can 
mimic orbital effects to some extent. This implies that modeling light curves 
of binary-lens events not considering orbital effects can result in lens parallaxes 
that are substantially different from actual values and thus wrong determinations 
of physical lens parameters. This demonstrates the importance of routine consideration 
of orbital effects in interpreting light curves of binary-lens events.
It is found that the lens of OGLE-2006-BLG-277 is a binary composed of a low-mass 
star and a brown dwarf companion.
\end{abstract}


\section{INTRODUCTION}

Progress in gravitational microlensing experiments for the last two decades has
enabled a great increase in the number of event detections from tens of events 
per year at the early stage to several thousands per year in current experiments. 
Among discovered lensing events, an important portion are produced by lenses
composed of two masses \citep{mao91}.

One reason why binary-lens events are important is that these events provide a major channel 
to determine the physical parameters of lenses. For the determination of the lens 
parameters from observed lensing light curves, it is required to simultaneously 
measure the lens parallax $\pi_{\rm E}$ and the angular Einstein radius $\theta_{\rm E}$. 
The lens parallax is measured from long-term deviations in lensing light curves 
caused by the positional change of an observer induced by the orbital motion of 
the Earth around the Sun: parallax effect \citep{gould92}. On the other hand, 
the Einstein radius is measured from deviations in lensing light curves affected 
by the finite size of a source star: finite-source effect \citep{gould94,witt94}. 
With the measured values of $\pi_{\rm E}$ and $\theta_{\rm E}$, the mass and 
distance to the lens are determined respectively by 
$M_{\rm tot}=\theta_{\rm E}/(\kappa\pi_{\rm E})$ and  
$D_{\rm L}={\rm AU}/(\pi_{\rm E}\theta_{\rm E}+\pi_{\rm S})$, 
where $\kappa=4G/(c^{2}{\rm AU})$, ${\rm AU}$ is an Astronomical Unit, 
$\pi_{\rm S}={\rm AU}/D_{\rm S}$, and $D_{\rm S}$ is the distance to the lensed star 
\citep{gould92,gould06}. For single-lens events, the chance to measure $\theta_{\rm E}$ 
is very low because finite-source effects occur only for very rare events with 
extremely high magnifications in which the lens passes over the surface of 
the source star, e.g., \citet{choi12}. By contrast, the chance to measure 
$\theta_{\rm E}$ is high for binary-lens events because most of these events 
involve source stars' caustic crossings or approaches during which 
finite-source effects are important. As a result, the majority of gravitational 
lenses with measured physical parameters are binaries.

It is known that changes of lens positions caused by the orbital motion of 
a binary lens can induce long-term deviations in lensing light curves, 
similar to deviations induced by parallax effects. Since this was first detected  
for the event MACHO-97-BLG-41 \citep{bennett99,albrow00,jung13}, 
orbital effects have been considered for more binary-lens events, e.g., 
\citet{an02,jaroszynski05,skowron11,shin11}. However, analyses have been carried 
out only for a limited number of events. An important obstacle 
of orbital analyses is the heavy computation required to consider the time variation 
of the caustic morphology caused by the orbit-induced changes of the binary 
separation and orientation. As a result, routine orbital analyses for general 
binary-lens events became possible very recently after being able to utilize 
efficient modeling software and powerful computing resources.

Considering orbital effects is important for accurate determinations of physical 
lens parameters. Since orbital and parallax effects induce similar long-term deviations, 
it might be that orbital effects can be mimicked by parallax effects. Then, 
if only parallax effects are considered for events affected by orbital effects, 
the determined physical parameters would be different from their true values. 
In this work, we demonstrate the importance of considering orbital effects 
by presenting analyses of two binary-lens events. 

% Table 1 ----------------------------------------------------
\begin{deluxetable}{ll}
\tablecaption{Telescopes\label{table:one}}
%\tablecolumns{12}
\tablewidth{0pt}
\tablehead{
\multicolumn{1}{l}{Event} & 
\multicolumn{1}{l}{Telescopes}
%\colhead{Event} & 
%\colhead{Telescopes} 
} 
\startdata
OGLE-2006-BLG-277   & OGLE, 1.3 m Warsaw, LCO, Chile       \\
                    & $\mu$FUN, 1.3 m SMARTS, CTIO, Chile  \\
                    & PLANET, 1.5 m Boyden, South Africa   \\
                    & PLANET, 1.0 m Canopus, Australia     \\
                    & PLANET, 0.6 m Perth, Australia       \\
                    & PLANET, 1.54 m Danish, Chile         \\
                    & RoboNet, 2.0 m LT, La Palma, Spain   \\
\hline
OGLE-2012-BLG-0031  & OGLE, 1.3 m Warsaw, LCO, Chile        \\
                    & $\mu$FUN, 1.3 m SMARTS, CTIO, Chile   \\
                    & $\mu$FUN, 0.36 m Turitea, New Zealand \\
                    & $\mu$FUN, 0.36 m KKO, South Africa    \\
                    & RoboNet, 2.0 m FTS, Australia         \\
                    & Robonet, 2.0 m LT, La Palma, Spain
\enddata
\tablecomments{LCO: Las Campanas Observatory,
CTIO: Cerro Tololo Inter-American Observatory,
KKO: Klein Karoo Observatory,
LT: Liverpool Telescope,
FTS: Faulkes Telescope South}
\end{deluxetable}
% -------------------------------------------------------------

\section{Observation}

The events analyzed in this work are OGLE-2006-BLG-277 and OGLE-2012-BLG-0031. 
Both events occurred on stars toward the Galactic bulge field with 
equatorial coordinates $(\alpha,\delta)_{\rm J2000}=(18^{\rm h}01^{\rm m}
14^{\rm s}\hskip-2pt.84, -27^{\circ}48'36''\hskip-2pt.2)$, 
corresponding to the Galactic coordinates $(l,b)=(2.71^\circ, -2.39^\circ)$, 
for OGLE-2006-BLG-277 and $(\alpha,\delta)_{\rm J2000}=(17^{\rm h}50^{\rm m}
50^{\rm s}\hskip-2pt.53, -29^{\circ}10'48''\hskip-2pt.8)$, corresponding to 
$(l,b)=(0.38^\circ, -1.10^\circ)$, for OGLE-2012-BLG-0031. 
The events were discovered from survey observations conducted by 
the Optical Gravitational Lensing Experiment \citep[OGLE:][]{udalski03}.
In addition to the survey observation, the events were additionally 
observed by follow-up groups including the Probing Lensing Anomalies 
NETwork \citep[PLANET:][]{albrow98}, the Microlensing Follow-Up Network 
\citep[$\mu$FUN:][]{gould06}, and the RoboNet \citep{tsapras09} groups. 
In Table~\ref{table:one}, we list survey and follow-up groups who participated in 
observations of the individual events along with the telescopes they employed as well as 
their locations. We note that the event OGLE-2006-BLG-277 was previously analyzed 
by \citet{jaroszynski10}, but the analysis was 
based on only OGLE data. We, therefore, reanalyze the event based on all 
combined data considering higher-order effects.

Reductions of data were done using photometry codes developed by the individual 
groups, mostly based on difference image analysis \citep{alard98,wozniak01,bramich08,albrow09}. 
In order to use data sets acquired from different observatories, 
we readjust the error bars. For this, we first add a quadratic error term 
so that the cumulative distribution of $\chi^{2}$ ordered by magnifications 
is approximately linear in data counts, and then 
rescale errors so that $\chi^{2}$ per degree of freedom $(\chi^{2}/{\rm dof})$ 
becomes unity. 

% Figure 1 ----------------------------------------------------
\begin{figure}[ht]
\epsscale{1.15}
\plotone{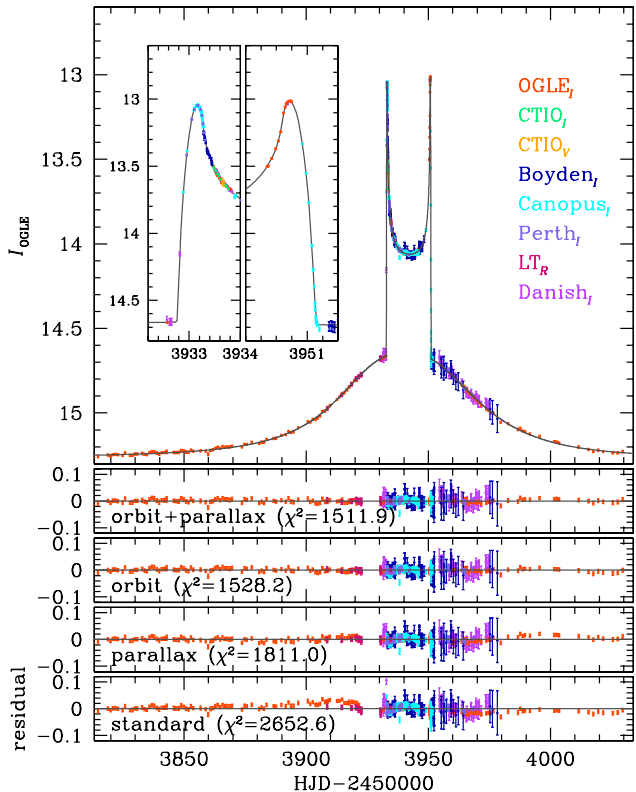}
\caption{\label{fig:one}
Light curve of OGLE-2006-BLG-277. In the legends indicating observatories, 
the subscript of each observatory denotes the passband. The subscript 
``$N$'' denotes that no filter is used.
The insets in the upper panel show the enlargement of the caustic-crossing 
parts of the light curve. The lower four panels show the residuals of data 
from four different models.
}\end{figure}
% -------------------------------------------------------------

In Figures~\ref{fig:one} and~\ref{fig:two}, we present the light curves of 
the individual events. Both light curves exhibit sharp spikes that are 
characteristic features of caustic-crossing binary-lens events. 
The spikes occur in pairs because the caustic forms a closed curve. 
Usually, the inner region between two spikes has a ``U''-shape trough as 
is in OGLE-2006-BLG-277. For OGLE-2012-BLG-0031, the inner region exhibits 
a complex pattern. Such a pattern can be produced when the source trajectory 
runs approximately tangent to the fold of a caustic.

\section{Modeling}

\subsection{Standard Model}

% Figure 2 ----------------------------------------------------
\begin{figure}[ht]
\epsscale{1.15}
\plotone{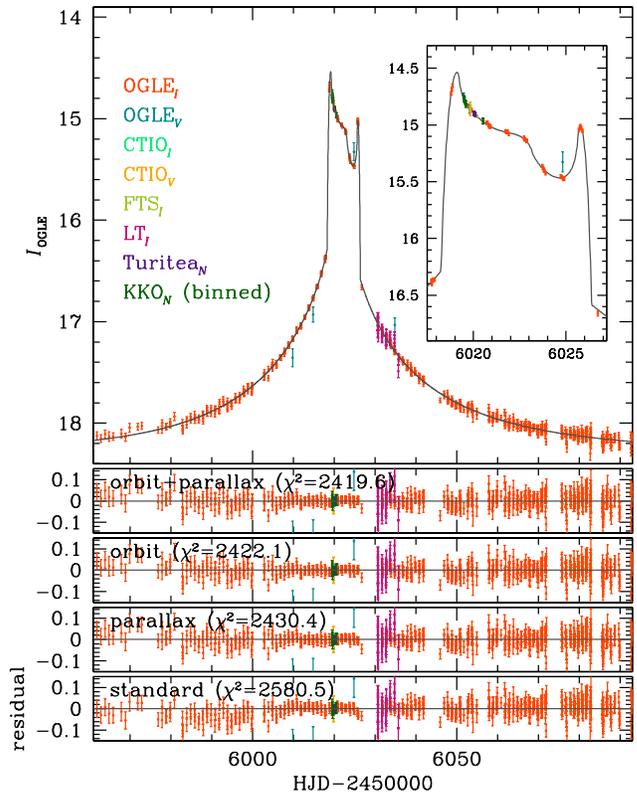}
\caption{\label{fig:two}
Light curve of OGLE-2012-BLG-0031.
Notations are the same as those in Fig. \ref{fig:one}.
}\end{figure}
% -------------------------------------------------------------

Knowing that the events were produced by binary lenses, we conduct 
modeling of the observed light curves. Basic description of a binary-lens 
event requires 7 lensing parameters. Among them, the first three parameters 
describe the lens-source approach. These parameters include the time of the closest 
approach of the source to a reference position\footnotemark[38]
% footnote -----------------------------------------------------------
\footnotetext[38]{
For a binary lens with a projected separation less than the Einstein radius, 
$s<1$, we set the reference position of the lens as the center of mass of 
the binary lens. For a binary with a separation greater than the Einstein radius, 
$s>1$, on the other hand, we set the reference as the photocenter that is 
located at a position with an offset $q/[s(1+q)]$ from the middle position 
between the two lens components.} 
% --------------------------------------------------------------------
of the binary lens, $t_{0}$, the lens-source separation at $t_{0}$ 
in units of the Einstein radius, $u_{0}$, and the time required for 
the source to cross the Einstein radius, $t_{\rm E}$ (Einstein time scale). 
The Einstein ring represents the source image for an exact lens-source 
alignment and its radius $\theta_{\rm E}$ is usually used as the length 
scale of lensing phenomena. The Einstein radius is related to the physical 
lens parameters by $\theta_{\rm E}=(\kappa{M}\pi_{\rm rel})^{1/2}$, 
where $M$ is the mass of the lens and 
$\pi_{\rm rel}={\rm AU}(D^{-1}_{\rm L}-D^{-1}_{\rm S})$ is the relative 
lens-source parallax. Another three lensing parameters 
describe the binary lens. These parameters include the projected separation, $s$ 
(in units of $\theta_{\rm E}$), the mass ratio between the binary lens 
components, $q$, and the angle between the source trajectory and the binary 
axis, $\alpha$ (source trajectory angle). The last parameter is the normalized source radius 
$\rho_{\ast}=\theta_{\ast}/\theta_{\rm E}$, where $\theta_{\ast}$ is the 
angular source radius. This parameter is needed to describe the parts of 
light curves affected by finite-source effects, which are important 
when a source star crosses over or approaches close to caustics formed by 
a binary lens.

In modeling the light curves based on the standard lensing parameters 
(standard model), searches for best-fit solutions 
have been done in two steps. In the first step, we identify local solutions 
by inspecting $\chi^{2}$ distributions in the parameter space. For this, 
we use both grid search and downhill approach. We choose ($s, q, \alpha$) 
as grid parameters because lensing magnifications can vary dramatically with 
small changes of these parameters. By contrast, lensing 
magnifications vary smoothly with changes of the other parameters, and 
thus we search for the solutions of these parameters by minimizing $\chi^{2}$ 
using a downhill approach.  We use the Markov Chain Monte Carlo (MCMC) 
method for the $\chi^{2}$ minimization. In the second step, we refine 
the lensing parameters for the individual local solutions by allowing 
all parameters to vary. Then, the best-fit solution is obtained by comparing 
$\chi^{2}$ values of the individual local solutions. We estimate 
the uncertainties of the lensing parameters based on the distributions 
of the parameters obtained from the MCMC chain of solutions.

% Table 2 ----------------------------------------------------
\begin{deluxetable*}{lrrrrrrrrrrrr}
\tablewidth{0pt}
\tablecaption{Model Parameters\label{table:two}}
\tablehead{
%\multicolumn{1}{l}{Event} & 
\multicolumn{1}{l}{Model} &
\multicolumn{1}{c}{$\chi^{2}/{\rm dof}$} & 
\multicolumn{1}{c}{$t_{0}$} & 
\multicolumn{1}{c}{$u_{0}$} & 
\multicolumn{1}{c}{$t_{\rm E}$} & 
\multicolumn{1}{c}{$s$} &
\multicolumn{1}{c}{$q$} &
\multicolumn{1}{c}{$\alpha$} &
\multicolumn{1}{c}{$\rho_{\ast}$} &
\multicolumn{1}{c}{$\pi_{{\rm E},N}$} &
\multicolumn{1}{c}{$\pi_{{\rm E},E}$} &
\multicolumn{1}{c}{$ds/dt$} &
\multicolumn{1}{c}{$d\alpha/dt$} \\
\multicolumn{1}{l}{} &
\multicolumn{1}{c}{} & 
\multicolumn{1}{c}{(HJD')} & 
\multicolumn{1}{c}{} & 
\multicolumn{1}{c}{(days)} & 
\multicolumn{1}{c}{} &
\multicolumn{1}{c}{} &
\multicolumn{1}{c}{} &
\multicolumn{1}{c}{($10^{-3}$)} &
\multicolumn{1}{c}{} &
\multicolumn{1}{c}{} &
\multicolumn{1}{c}{(${\rm yr}^{-1}$)} &
\multicolumn{1}{c}{(${\rm yr}^{-1}$)}
} 
\startdata
OGLE-2006-BLG-277       &            &            &           &            &            &            &           &                           &                           &                           &                           \\
Standard       & 2652.6 & 3941.620   & 0.157      & 39.13     & 1.374      & 2.600      & 1.477      & 5.83      & \multirow{2}{*}{$\cdots$} & \multirow{2}{*}{$\cdots$} & \multirow{2}{*}{$\cdots$} & \multirow{2}{*}{$\cdots$} \\
               & /1499  & $\pm$0.020 & $\pm$0.002 & $\pm$0.08 & $\pm$0.001 & $\pm$0.037 & $\pm$0.003 & $\pm$0.02 &                           &                           &                           &                           \\
Parallax only  & 1811.0 & 3941.723   & 0.169      & 39.30     & 1.371      & 2.512      & 1.485      & 5.90      & 0.45                      & 0.54                      & \multirow{2}{*}{$\cdots$} & \multirow{2}{*}{$\cdots$} \\
               & /1497  & $\pm$0.025 & $\pm$0.003 & $\pm$0.08 & $\pm$0.001 & $\pm$0.035 & $\pm$0.003 & $\pm$0.02 & $\pm$0.07                 & $\pm$0.01                 &                           &                           \\
Orbit only     & 1528.2 & 3943.066   & 0.170      & 38.78     & 1.347      & 2.033      & -1.485     & 5.98      & \multirow{2}{*}{$\cdots$} & \multirow{2}{*}{$\cdots$} & 0.73                      & -0.33                     \\
               & /1497  & $\pm$0.031 & $\pm$0.005 & $\pm$0.07 & $\pm$0.001 & $\pm$0.030 & $\pm$0.005 & $\pm$0.02 &                           &                           & $\pm$0.02                 & $\pm$0.11                 \\
Orbit+Parallax & 1511.9 & 3943.071   & -0.168     & 37.90     & 1.348      & 1.981      & 1.457      & 6.03      & 1.13                      & -0.05                     & 0.69                      & 1.21                      \\
               & /1495  & $\pm$0.031 & $\pm$0.005 & $\pm$0.13 & $\pm$0.001 & $\pm$0.030 & $\pm$0.006 & $\pm$0.02 & $\pm$0.16                 & $\pm$0.04                 & $\pm$0.03                 & $\pm$0.22                 \\ 
\tableline
OGLE-2012-BLG-0031      &            &            &           &            &            &            &           &                           &                           &                           &                           \\
Standard       & 2580.5 & 6022.532   & 0.046      & 59.17     & 0.477      & 0.294      & 0.800      & 5.48      & \multirow{2}{*}{$\cdots$} & \multirow{2}{*}{$\cdots$} & \multirow{2}{*}{$\cdots$} & \multirow{2}{*}{$\cdots$} \\
               & /2403  & $\pm$0.042 & $\pm$0.001 & $\pm$0.59 & $\pm$0.003 & $\pm$0.010 & $\pm$0.009 & $\pm$0.11 &                           &                           &                           &                           \\
Parallax only  & 2430.4 & 6022.233   & -0.047     & 56.47     & 0.510      & 0.223      & -0.739     & 5.63      & -0.29                     & 0.10                      & \multirow{2}{*}{$\cdots$} & \multirow{2}{*}{$\cdots$} \\
               & /2405  & $\pm$0.043 & $\pm$0.001 & $\pm$0.66 & $\pm$0.003 & $\pm$0.008 & $\pm$0.009 & $\pm$0.11 & $\pm$0.08                 & $\pm$0.02                 &                           &                           \\
Orbit only     & 2422.1 & 6022.364   & 0.051      & 54.88     & 0.511      & 0.264      & 0.774      & 6.80      & \multirow{2}{*}{$\cdots$} & \multirow{2}{*}{$\cdots$} & 0.43                      & 3.63                      \\
               & /2405  & $\pm$0.042 & $\pm$0.001 & $\pm$0.68 & $\pm$0.003 & $\pm$0.011 & $\pm$0.009 & $\pm$0.19 &                           &                           & $\pm$0.08                 & $\pm$0.20                 \\
Orbit+Parallax & 2419.6 & 6022.350   & -0.051     & 54.58     & 0.511      & 0.268      & -0.773     & 6.81      & -0.09                     & 0.03                      & 0.47                      & -2.98                     \\
               & /2407  & $\pm$0.042 & $\pm$0.001 & $\pm$0.77 & $\pm$0.003 & $\pm$0.010 & $\pm$0.009 & $\pm$0.21 & $\pm$0.13                 & $\pm$0.02                 & $\pm$0.07                 & $\pm$0.39 
\enddata
\tablecomments{HJD'=HJD-2,450,000}
\end{deluxetable*}
% -------------------------------------------------------------

For magnification computations affected by finite-source effects, 
we use the ``map-making method'' developed by \citet{dong06}. In this method, 
a map of rays for a given binary lens with a separation $s$ and a mass ratio $q$ 
is constructed by using the inverse ray-shooting technique 
\citep{schneider86,kayser86,wambsganss97}.
In this technique, rays are uniformly shot from the image plane, 
bent according to the lens equation, and land on the source plane. 
The lens equation for a binary lens is represented by
\begin{equation}
\zeta=z-{m_{1}\over \overline{z}-\overline{z}_{\rm L,1}}
-{m_{2}\over \overline{z}-\overline{z}_{\rm L,2}},
\label{eq1}
\end{equation}
where $m_{1}={1}/(1+q)$ and $m_{2}=qm_{1}$ are the mass fractions of 
the individual binary lens components, $\zeta=\xi+i\eta$, $z=x+iy$, and 
$z_{{\rm L},i}=x_{{\rm L},i}+iy_{{\rm L},i}$ denote the positions of the source, 
image, and lens expressed in complex notions, respectively, and $\overline{z}$ denotes 
the complex conjugate of $z$. With the constructed map, the finite-source magnification for 
a given position of a source with a normalized radius $\rho_{\ast}$ is 
computed as the ratio of the number density of rays within the source 
to that on the image plane. This method saves computation time by 
enabling to produce many light curves resulting from various source trajectories 
based on a single map. In addition, the method enables to speed 
up computation by allotting computation into multiple CPUs. We further 
accelerate computation by using semi-analytic hexadecapole approximation 
\citep{pejcha09,gould08} for finite magnification computations.

In our finite-source computations, we consider the limb-darkening effect 
of the source star by modeling the surface brightness profile as 
$S_{\lambda}\propto1-\Gamma_{\lambda}(1-3\cos \phi /2)$, where 
$\Gamma_{\lambda}$ is the linear limb-darkening coefficient, $\lambda$ 
is the passband, and $\phi$ denotes the angle between the line of sight 
toward the source star and the normal to the source surface. 
The limb-darkening coefficients are adopted from \citet{claret00} 
considering the source type that is determined based on the source locations 
in the color-magnitude diagrams. 
We find that the source star of OGLE-2006-BLG-277 is a K-type giant star.
For OGLE-2012-BLG-0031, the lensed star is located in a very reddened region, 
causing difficulties in precisely characterizing the star based on its 
color and brightness. Nevertheless, it is found that the source is a giant. 
The adopted coefficients are $\Gamma_{V}=0.74$, $\Gamma_{R}=0.64$ and $\Gamma_{I}=0.53$ for both events.
For data sets obtained without any filter, we choose a mean value of the $R$ and $I$ band coefficients, 
i.e., $\Gamma_{N}=(\Gamma_{R}+\Gamma_{I})/2$, where the subscript ``$N$''
denotes that no filter is used.

In Table~\ref{table:two}, we list the best-fit solutions of the lensing 
parameters obtained from standard modeling for the individual events. 
In Figures~\ref{fig:one} and~\ref{fig:two}, we also present the residuals 
from the fits. It is found that even though the fits basically describe 
the main features of the observed light curves, there exist important 
residuals that last throughout both events.

\subsection{Higher-order Effects}

Long-term residuals from the standard models 
suggest that one needs to consider higher-order effects in order to better 
describe the lensing light curves. Since it is known that 
such long-term residuals are caused by the parallax and/or lens orbital effects, 
we conduct additional modeling considering both higher-order effects. 

To describe parallax effects, it is necessary to include two 
parameters $\pi_{{\rm E},N}$ and $\pi_{{\rm E},E}$, which represent 
the two components of the lens parallax vector $\pivec_{\rm E}$ projected onto 
the sky along the north and east equatorial coordinates, respectively. 
The magnitude of the parallax vector, ${\pi}_{\rm E}=(\pi_{{\rm E},N}^{2}
+\pi_{{\rm E},E}^{2})^{1/2}$, corresponds to the relative lens-source parallax 
scaled to the Einstein radius of the lens, i.e. $\pi_{\rm E}=\pi_{\rm rel}/\theta_{\rm E}$ 
\citep{gould04}. The direction of the parallax vector corresponds to 
the relative lens-source motion in the frame of the Earth at a reference time 
of the event. In our modeling, we use $t_{0}$ as the reference time. 
Parallax effects cause the source motion relative to the lens to deviate
from rectilinear.

To first order approximation, the lens orbital motion is described by 
two parameters $ds/dt$ and $d\alpha/dt$, which
represent the change rates of the normalized binary separation and 
the source trajectory angle, respectively \citep{albrow00,an02}.
In addition to causing the relative lens-source motion to deviate 
from rectilinear, the orbital effect causes further deviation in 
lensing light curves by deforming the caustic over the course of 
the event due to the change of the binary separation.

% Figure 3 ----------------------------------------------------
\begin{figure}[ht]
\epsscale{1.15}
\plotone{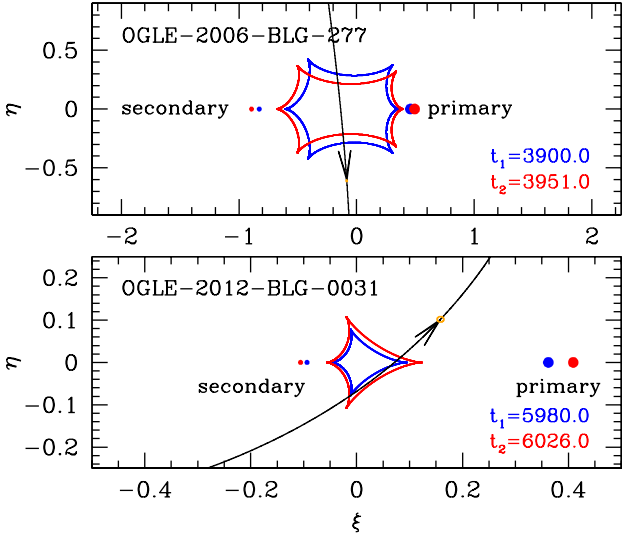}
\caption{\label{fig:three}
Geometry of the best-fit models for OGLE-2006-BLG-277 (upper panel) and 
OGLE-2012-BLG-0031 (lower panel). Small dots and closed 
solid curves represent the lens positions and caustics at two different
times $t_1$ and $t_2$. Black solid curves with arrows represent 
source trajectories. The size of the small empty circle at the tip of 
the arrow of each source trajectory represents the source size.
The abscissa and ordinate are parallel with and perpendicular to the 
binary axis, respectively. All lengths are normalized by the Einstein 
ring radius.
}\end{figure}
% -------------------------------------------------------------

In Table~\ref{table:two}, we list the results of modeling considering 
the higher-order effects. For each event, we conduct 3 sets of additional 
modeling in which the parallax effect and orbital effect are considered 
separately (``parallax only'' and ``orbital only'') and both effects are 
simultaneously considered (``orbit+parallax''). In the lower panels of 
Figures~\ref{fig:one} and~\ref{fig:two}, we present the residuals of 
the individual models. In Figure~\ref{fig:three}, we present the geometry 
of the lens systems of the best-fit solutions, where the source trajectory 
with respect to the lens components and the resulting caustics are shown. 
We note that the relative lens positions and caustics vary in time due to 
the orbital motion of the lens and thus we mark the positions 
at two different moments.

For both events, we find that the dominant second-order effect is 
the lens orbital motion. The dominance of the orbital effect is evidenced 
by the fact that the models considering only the orbital effect result 
in fits as good as those considering both the parallax and orbital effects. 
It is found that the consideration of orbital effects improves the fits 
by $\Delta \chi^{2}=1124.4$ and $158.4$ 
compared to the standard models of OGLE-2006-BLG-277 and OGLE-2012-BLG-0031, 
respectively. On the other hand, the improvements by additionally considering 
the parallax effect are merely $\Delta \chi^{2}=16.3$ and $2.5$ for the 
individual events. 

% Figure 4 ----------------------------------------------------
\begin{figure}[ht]
\epsscale{1.15}
\plotone{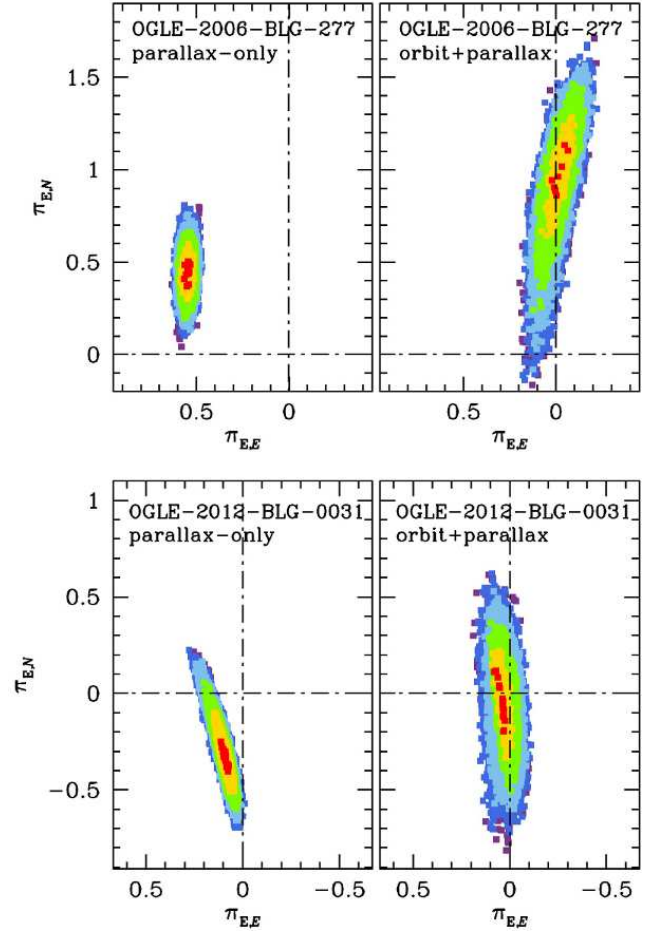}
\caption{\label{fig:four}
Distributions of $\chi^{2}$ in the space of the parallax parameters
$\pi_{{\rm E},E}$ and $\pi_{{\rm E},N}$ for OGLE-2006-BLG-277 (upper panels) 
and OGLE-2012-BLG-0031 (lower panels). For each event, the distribution in
the left panel is obtained from modeling considering only the parallax
effect, while the distribution in the right panel is constructed by considering
both the orbital and parallax effects. Different contours correspond to 
$\Delta\chi^{2}<1$ (red), $4$ (yellow), $9$ (green),
$16$ (sky blue), $25$ (blue), and $36$ (purple), respectively.
}\end{figure}
% -------------------------------------------------------------

To be noted is that parallax effects can mimic orbital effects to some extent 
for both events. We find that the improvements of the fits by the parallax 
effect are $\Delta \chi^{2}=841.6$ (cf. $\Delta \chi^{2}=1124.4$ improvement 
by the orbital effect) and $150.1$ (cf. $\Delta \chi^{2}=158.4$ by 
the orbital effect) for OGLE-2006-BLG-277 and OGLE-2012-BLG-0031, respectively. 
In addition, the values of the lens parallax determined without considering orbital effects 
substantially differ from those determined by considering orbital effects. This can be seen in 
Figure~\ref{fig:four} where we present $\chi^{2}$ distributions in the space of the 
parallax parameters. For OGLE-2006-BLG-277, the measured lens parallax is $\pi_{\rm E}=1.13\pm0.16$ 
when both parallax and orbital effects are considered, while $\pi_{\rm E}=0.70\pm0.05$ 
when only the parallax effect is considered. For OGLE-2012-BLG-0031, modeling considering 
only parallax effects results in a lens parallax $\pi_{\rm E}=0.31\pm0.08$, while 
the lens parallax is consistent with zero in 3$\sigma$ level in the model 
considering additional orbital effects.
These facts imply that orbital effects can masquerade as parallax effects 
and thus lens parallax values measured based on modeling not considering orbital 
effects can result in wrong values. This leads to wrong determinations of
physical lens parameters because masses and distances to lenses are determined 
from measured values of the lens parallax.

It was pointed out by \citet{batista11} and \citet{skowron11} that the parallax component
perpendicular to the relative lens-source motion, $\pi_{{\rm E},\bot}$, 
is strongly correlated with the orbital parameter $d\alpha/dt$, causing a 
degeneracy between $\pi_{{\rm E},\bot}$ and $d\alpha/dt$. They argued that 
this degeneracy occurs because the lens-source motion in the direction 
perpendicular to the Sun-Earth axis induces deviations in lensing light curves 
similar to those induced by the rotation of the binary-lens axis. For both 
events OGLE-2006-BLG-277 and OGLE-2012-BLG-0031, the direction of the relative 
lens-source motion is similar to east-west direction, and thus $\pi_{{\rm E},\bot} 
\sim \pi_{{\rm E},N}$. According to this degeneracy, the lens parallax vectors 
estimated by the ``parallax only'' and the ``orbit+parallax'' models should 
result in similar values of $\pi_{{\rm E},E}$, while values of $\pi_{{\rm E},N}$ 
can be widely different. However, both the events analyzed in this work do not 
conform to the previous prediction. This implies that the parallax-orbit degeneracy 
is much more complex than previously thought, and thus it is essential to study 
the degeneracy in all cases where higher-order effects are detected.

We determine the physical lens parameters based on the best-fit solutions (orbit+parallax models). 
For this, we first determine the Einstein radius. 
The Einstein radius is determined by $\theta_{\rm E}=\theta_\ast/\rho_\ast$, 
where the normalized source radius $\rho_\ast$ is measured from the modeling 
and the angular stellar radius is determined based on the source type. 
The measured Einstein radius of the lens of OGLE-2006-BLG-277 is 
is $\theta_{\rm E}=1.35\pm0.11$ ${\rm mas}$. This corresponds to the relative lens-source 
proper motion $\mu=\theta_{\rm E}/t_{\rm E}=13.0\pm1.0$ ${\rm mas}$ ${\rm yr^{-1}}$. 
With the measured mass ratio between the lens components, then the masses of the individual 
lens components are $M_{1}=M_{\rm tot}/(1+q)=0.049\pm0.014$ $M_{\odot}$ and $M_{2}=qM_{\rm tot}/(1+q)
=0.097\pm0.027$ $M_{\odot}$, respectively. Therefore, the lens is composed of a low-mass star 
and a brown dwarf. The distance to the lens is $D_{\rm L}=0.60\pm0.14$ ${\rm kpc}$. 
The close distance explains the relatively high proper motion ($13.0\pm1.0$ ${\rm mas}$ ${\rm yr^{-1}}$).
With the physical parameters combined with orbital parameters, we evaluate the ratio of tranverse 
kinetic to potential energy
\begin{equation}
\left( {\rm KE \over \rm PE} \right)_{\bot} = {(r_{\bot}/{\rm AU})^{2} \over 8{\pi}^{2}(M_{\rm tot}/M_{\odot})}
\left[ \left( {1 \over s}{ds \over dt} \right)^{2} + \left( { d\alpha \over dt} \right)^{2} \right],
\label{eq2}
\end{equation}
where $r_{\bot}$ denotes the projected binary separation \citep{dong09}.
The ratio should obey $\rm (KE/PE)_{\bot} \leq KE/PE < 1$ for kinetically stable binary orbit. 
We find $\rm (KE/PE)_{\bot}=0.20\pm0.04$.
For OGLE-2012-BLG-0031, it is difficult to determine the physical lens 
parameters not only because the source type is uncertain but also because the lens parallax 
is consistent with zero.

\section{Summary and Conclusion}

We analyzed two binary-lens events OGLE-2006-BLG-277 
and OGLE-2012-BLG-0031 for which the light curves exhibit significant 
residuals from standard binary-lens models. From modeling considering 
higher-order effects, we found that the residuals were greatly removed 
by considering the effect of the lens orbital motion. We also found that 
parallax effects could mimic orbital effects to some extent and the 
parallax values measured not considering the orbital effect could result 
in dramatically different value from true values, and thus wrong determinations 
of physical lens parameters. We also found that the lens of OGLE-2006-BLG-277 
was a binary composed of a low-mass star and a brown dwarf companion.

\acknowledgments
Work by CH was supported by Creative Research Initiative Program
(2009-0081561) of National Research Foundation of Korea.
AG was supported by NSF grant AST 1103471.
The OGLE project has received funding from the European Research Council 
under the European Community's Seventh Framework Programme (FP7/2007-2013) 
/ ERC grant agreement no. 246678 to AU.
S.D. was supported through a Ralgh E. and Doris M. Hansmann Membership at
the IAS and NSF grant AST-0807444.
DMB, MD, MH, RAS and YT would like to thank the Qatar Foundation for 
support from QNRF grant NPRP-09-476-1-078. 
Dr. David Warren gave financial support to Mt. Canopus Observatory.

\end{document}